\documentclass[journal]{IEEEtran}
\makeatletter
\def\@copyrightspace{\relax}
\makeatother
\usepackage[utf8]{inputenc}

\usepackage{cite}
\usepackage{amsmath,amssymb,amsfonts}
\usepackage{mathtools}
\usepackage{graphicx}
\usepackage{caption}
\usepackage{textcomp}
\usepackage{xcolor}
\usepackage{comment}
\usepackage{tabularx,ragged2e,booktabs}
\usepackage{makecell}
\usepackage{multirow}
\usepackage[most]{tcolorbox}
\usepackage[justification=centering]{caption}
\usepackage[utf8]{inputenc}
\usepackage{longtable}
\usepackage{caption}
\usepackage{lipsum}% http://ctan.org/pkg/lipsum
\usepackage[english]{babel}
\usepackage[utf8]{inputenc}
\usepackage{algpseudocode} 
\usepackage{booktabs}
\renewcommand{\baselinestretch}{0.995}
\def\BibTeX{{\rm B\kern-.05em{\sc i\kern-.025em b}\kern-.08em
    T\kern-.1667em\lower.7ex\hbox{E}\kern-.125emX}}

\title{GF-Flush: A GF(2) Algebraic Attack on \\ Secure Scan Chains
\\
}    
\makeatletter
\newcommand{\linebreakand}{%
  \end{@IEEEauthorhalign}
  \hfill\mbox{}\par
  \mbox{}\hfill\begin{@IEEEauthorhalign}
}
\makeatother
\author{Dake~Chen,~\IEEEmembership{Member,~IEEE,} Chunxiao~Lin,
Peter~A.~Beerel,~\IEEEmembership{Senior Member,~IEEE}}
    
\begin{document}

\maketitle

\begin{abstract}
Scan chains provide increased controllability and observability for testing digital circuits. The increased testability, however, can also be a source of information leakage for sensitive designs. The state-of-the-art defenses to secure scan chains apply dynamic keys to pseudo-randomly invert the scan vectors. In this paper, we pinpoint an algebraic vulnerability of these dynamic defenses that involves creating and solving a system of linear equations over the finite field GF(2). In particular, we propose a novel GF(2)-based flush attack that breaks even the most rigorous version of state-of-the-art dynamic defenses. 
Our experimental results demonstrate that our attack recovers the key as long as 500 bits in less than 7 seconds, the attack times are about one hundredth of state-of-the-art SAT based attacks on the same defenses. 
We then demonstrate how our attacks can be extended to scan chains compressed with Multiple-Input Signature Registers (MISRs).
%a fraction of a second.
\end{abstract}

\section{Introduction}
\label{sec:intro}
% placeholder
% placeholder=>summarize normal SAT attack
% placeholder=>summarize static scan obfuscation techniques

% 3 dynamic scan obfuscation papers
% scan sat+nyu attack
The decentralized supply chain of modern integrated circuit (IC) design and manufacturing raises significant concern related to threats that include intellectual property (IP) piracy~\cite{roy2010ending} and Trojan insertion~\cite{tehranipoor2010survey}. For many designs, the scan chain used in manufacturing testing presents a significant threat vector as it provides extensive controllability and observability of chip internals to the attacker~\cite{1677712,5419848, DaRolt2012ANS}.

The state of the art defenses involve applying dynamic keys to obfuscate the scan chain \cite{8709792,8105900,Rahman2019DynamicallyOS}. They leverage a linear feedback shift register (LFSR) that controls XOR gates along the scan chain to psuedo-randomly invert the scan chain sequence.
%in every cycle to encrypt locking gates, XOR or MUX, on the scan chain. 
The psuedo-random sequence is dependent on the seed of the LFSR which must 
remain secret to ensure security.
%Due to the property of the LFSR, all generated keys are based on a secret seed, thus the key to breaking these designs is to unveil this seed. 
Recently, some SAT based attacks \cite{8836102, 9116197} were proposed to unveil the seed by converting the scan flip-flops to psuedo input and outputs and thereby modeling the sequential circuit and LFSR as a combinational circuit that can be analyzed through well-known SAT attacks. The work \cite{Rahman2019DynamicallyOS} points out that this conversion from sequential to combinational logic increases the number of SAT literals and clauses, increasing the complexity and associated run-times of SAT attacks. 
%because the key is updated in every cycle, and this impact is exacerbated in long keys. As a consequence, as the increase of the key bits and size of the dynamic obfuscation designs, the complexity of the SAT based attacks increase exponentially, which makes the attack difficult.

In contrast, a simple flush and reset attack was proposed in \cite{4352010}. Here all flip-flops on scan chain are reset to 0 and the attack examines the initial sequence of scan out bits. Because the attacker can also reverse engineer the location of the locking gates, they are able to reveal the key input values from the scan out patterns. One recent dynamic obfuscation design \cite{8105900,Rahman2019DynamicallyOS} resists this reset attack by adding a shadow chain between the LFSR and scan chain. Due to the presence of the shadow chain which has the same length as LFSR, the initial scan out patterns remain zero and leak no information about the secret seed. %Besides, similar algebraic techniques which find the seed and characteristic polynomial of LFSR by analyzing the relationship between the seed and desired test patterns are utilized in EDA domain~\cite{658569,Wunderlich1987SelfTU}, but it has never been applied in hardware security.

%It is worth noting that the LFSR based security systems are easily susceptible to this attack because of the properties in LFSR~\cite{Algebraic1}. 
%Our attack can be categorized into this attack.

% in DOS, not only  scan locking, RLL for combinational locking, increase the complexity of Sat ATTACK
% scan sat + slides
% shadow schain

% Our attack only perform on scan chain and algebra derivation,  don't need to deal wtih RLL in combinational circuit, and focus on the scan lock
% break in seconds
% shadow chain does not work, because the A matrix which is the transition matrix of DOS is periodical

In this paper, we propose a more comprehensive flush attack based on GF(2) algebra that unveils the secret key of the dynamic scan locking defenses even when protected by a shadow chain. In contrast to SAT-based attacks~\cite{8836102, 9116197} which attack the scan chain coupled with locked combinational logic, our attack isolates the scan chain, enabling the use of more computationally scalable algebraic techniques used in crypto-analysis~\cite{ courtois2007algebraic}, including attacks on LFSRs  \cite{Courtois2003}, and  automatic test pattern generation \cite{658569,Wunderlich1987SelfTU}.
%(see e.g., \cite{Algebraic1, courtois2007algebraic,simmons2009algebraic,nakahara2009linear,cid2009block})
In particular, the attack involves solving a system of linear equations over the finite field GF(2) whose size scales linearly with the size of the key. We empirically validate that the complexity of our attack scales as a low-degree polynomial, recovering the key
%seed bits for LFSRs 
that is as long as 500 bits in less than 7 seconds. The  attack  times are  about  one  hundredth  of  state-of-the-art  SAT  based  attacks on  the  same  defense.

We further consider the case when the only access to the scan chain outputs is through test compression logic, such as a Multiple-Input Signature Registers (MISR). 
Because MISRs also consists of XOR gates and FFs they can be modeled, analyzed, and thus included in our attack.
To the best of our knowledge, this is the first attack on obfuscated scan chains that considers the impact of test compression logic.
%and circumvents the use of shadow chains \cite{8105900,Rahman2019DynamicallyOS}  
%designed to mitigate reset attacks. 
Although slower with MISRs, we demonstrate 
our attack times remain manageable.
%Besides, we also demonstrate our attack on ISCAS-89 benchmarks protected by dynamic scan locking techniques.
%In contrast, SAT-based attacks time out with LFSRs as short as 30 bits.

The remainder of this paper is organized as follows. Section~\ref{sec:backg} reviews the background leveraged in this paper. Section~\ref{sec:flushattack} describes the proposed attack. Section~\ref{sec: exp} details experimental results of our attack. Some conclusions and opportunities for future work are discussed in the last section.

\section{Background}
\label{sec:backg}

%\begin{figure*}
%    \centering
%    \includegraphics[width=12cm]{Generic_dos.jpg}
%    \caption{Basic structure of a dynamically secured scan chain}
%    \label{fig:DOS}
%\end{figure*}

\begin{figure*}
    \centering
    \includegraphics[width=18cm]{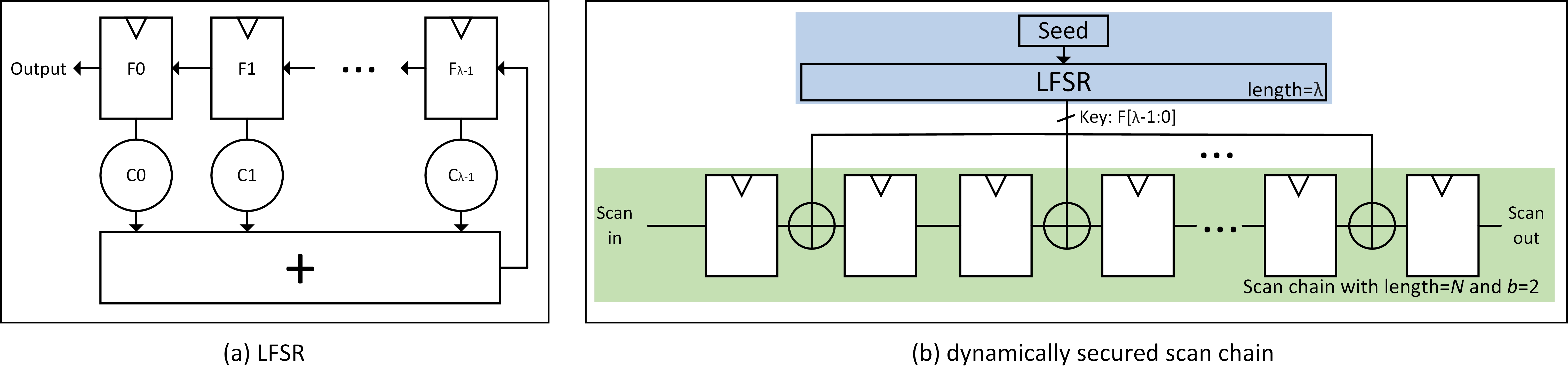}
    \caption{LFSR and basic structure of a dynamically secured scan chain}
    \label{fig:LFSR+DOS}
\end{figure*}

\subsection{A Linear Feedback  Shift  Register  (LFSR) }
% LFSR-technical document, all operations are in GF(2)
A Linear Feedback Shift Register (LFSR) is often used as pseudo-random number generator in many cryptographic and secure systems because of its lightweight, low overhead and high throughput~\cite{ShivaPrasad2019, 6119932}.
%

%\begin{figure}[hbt!]
%    \centering  
%    \includegraphics[width=7cm]{LFSR.jpg}
%    \caption{Generic structure of LFSR}
%    \label{fig:LFSR}
%\end{figure}
% In LFSR, all operations are in GF(2). 
The generic structure of an LFSR is shown in Figure~\ref{fig:LFSR+DOS}, where $\lambda$ denotes its length and the Binary values $c_0$ to $c_{\lambda-1}$ determine its feedback structure. The next state equation $f_i^{t+1}$ can be represented as
 \begin{align}
    %\begin{cases}
      f^{t+1}_{i} & =f^{t}_{i+1}, \text{ for $i\in [0,\lambda-1)$} \\
      f^{t+1}_{\lambda-1} & =\sum_{j=0}^{\lambda-1}c_j f^t_j 
 %   \end{cases}       
    \label{eq:lfsr1}
\end{align}
where $t$ and $t+1$ represent the current and next state, respectively, $f_i^t$ denotes the value of stage $i$ of LFSR at time $t$, and all operations are in GF(2).

The sequence generated by an LFSR is periodic and the period depends on the values of $c_i$ and the initial state, or \emph{seed} of the LFSR.
%and the period depends on the values of $c_i$. 
The maximum period of an LFSR of length $\lambda$ is $2^\lambda-1$ 
%because the state vector 0 is terminal 
\cite{WardlawLOT}. The sequences generated by LFSRs with maximum period are referred to as PN-sequences and these are desired for secure systems as they are more difficult to 
break than LFSRs with small periods.
%In the security systems which apply LFSR, larger period which achieve higher level of security is always desired.

\subsection{Dynamically Obfuscated Scan Chains}
\label{bkg:dos}

Due to the effectiveness of SAT-based attacks \cite{8836102} on static scan chain obfuscation techniques\cite{Karmakar2018EncryptFA}, state-of-the-art secure chains dynamically obscure scan chains using XORs that are driven by an LFSR \cite{8709792,8105900,Rahman2019DynamicallyOS} and psuedo-randomly invert the scan sequence.\footnote{MUXes can also be used to selectively invert the scan bit by muxing between the $Q$ and $Q_{bar}$ outputs of the scan FFs \cite{8709792}.} 
The basic structure of these schemes is shown in Figure~\ref{fig:LFSR+DOS}, 
where $\lambda$ represents the length of the LFSR and key, $N$ denotes the length of scan chain, and $b$ represents the spacing of locking gates throughout the chain. The most secure version of these methods updates the LFSR every clock cycle, applying new key bits to the scan locking gates every 
cycle.
%Furthermore, the two scan locking gates can be abstracted by using the same algebraic model in our attack.

% generic dynamic scan obfuscation structure- good figure on scansat paper
% DOSC structure-shadow layer-good summary on scansat paper
%The most secure version of DOSC in which the key updated rate is the same as scan clock frequency 

\subsection{MISR}

As the size of chips and number of scanned FFs increase, the latency and memory requirements to shift out and process their stored values during test grows.  For this reason test compression techniques, involving both a decompressor and compressor, have become an essential part of the design. The decompressor expands one scanned-in sequence into many parallel scan chain segments and the compressor compresses the outputs of many parallel scan segments into one. The most commonly used compressor is a Multiple-Input Signature Register (MISR)~\cite{620315} illustrated in Figure~\ref{fig: DOSwithMISR}, 

Because the MISR can prohibit direct access to the scan outputs, it has significant impact on all HW security attacks that rely on scan chain access, including previous SAT-based attacks~\cite{8836102, 9116197}.  Interestingly, as the MISR uses XOR gates that are commonly used to obfuscate combinational logic, one might think the MISR effectively encrypts the scan outputs. 

\subsection{Algebraic Analysis}

LFSRs are commonly used in built-in-self-test structures 
and algebraic analysis establishing the relationship between seed and outputs ~\cite{658569,Wunderlich1987SelfTU} has been used to find seeds and characteristic polynomials that lead to high test coverage. Moreover, algebraic cryptanalysis or algebraic attack~\cite{Courtois2003, courtois2007algebraic} has been widely used for attacking various ciphers. These attacks first find low degree equations to approximate the function of feedback shift registers (FSR) or algorithms based on their features, then leverage the XL algorithm~\cite{XLalgo2000} to solve the system of multivariate polynomial equations, thereby acquire the key bits.
These algebraic techniques, however, have never been applied in scan-chain locking.
%This type of attacks which target at algebraic vulnerability has never been used for attacking dynamic scan locking techniques. 
Considering all operations in the LFSR, scan-chain locking gates and MISR are effectively XOR operations, we hypothesize that an algebraic attack over GF(2) can be very efficient.

\section{GF-Flush: A GF(2) Algebraic Attack}
\label{sec:flushattack}
% The matrix representation of LFSR states reveals many properties of it.
\subsection{Algebraic Foundations of the Attack}
\begin{figure}[bt!]
    \centering  
    \includegraphics[width=7cm]{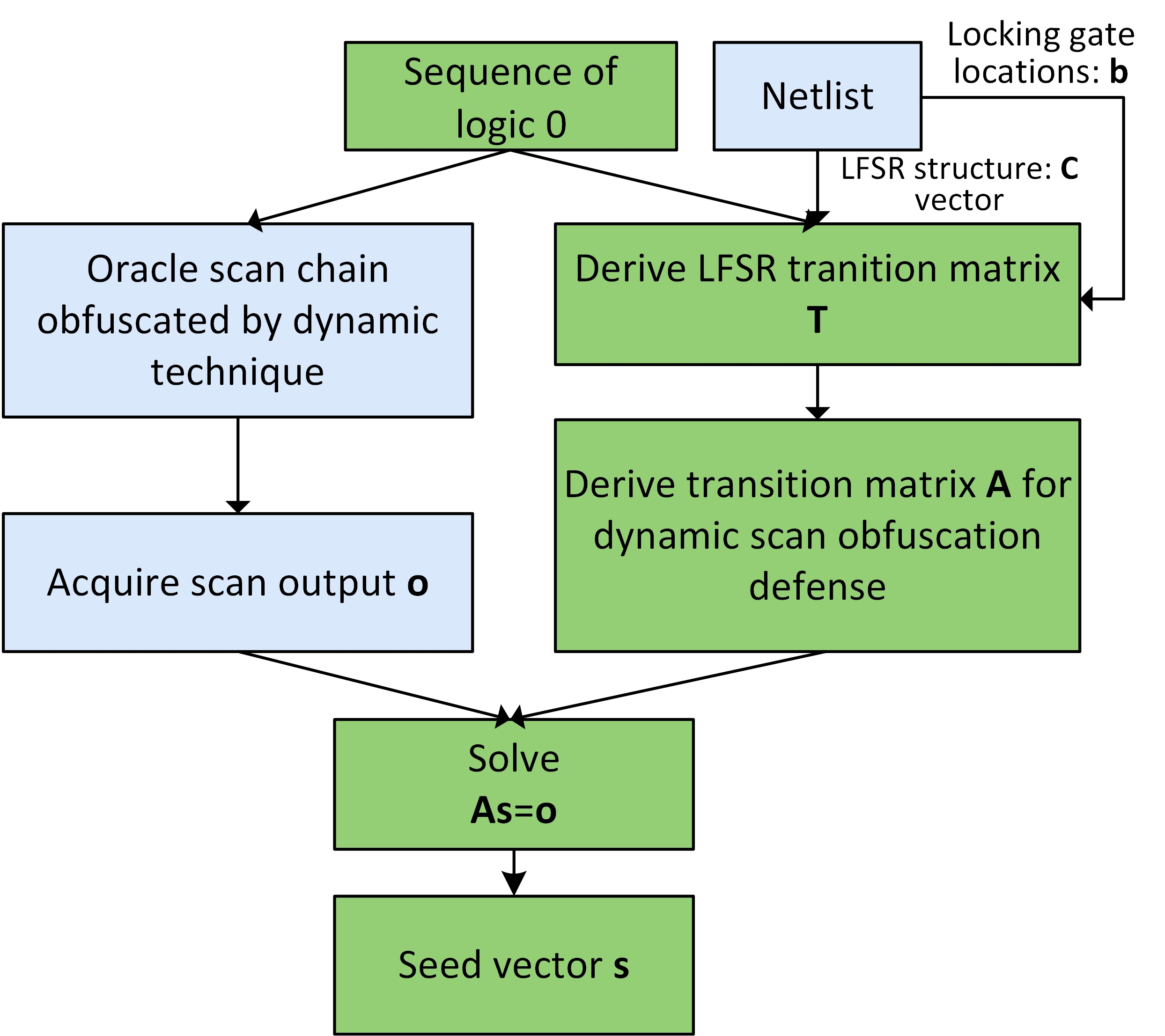}
    \caption{Flow of the proposed attack}
    \label{fig: attackflow}
\end{figure}

%Because all operations in both the LFSR and scan-chain are effectively XOR operations, we hypothesize that an algebraic attack over GF(2) can be very efficient. In particular, 
The basic flow of our proposed attack is illustrated in Figure~\ref{fig: attackflow}. 
Similar to previous attacks on the same defenses~\cite{9116197}, we assume that the netlist is reverse-engineered and thus the structural information about the LFSR $c_i$, the length of the scan chain $N$, and the location of XOR gates $b$ are known to the attacker. We also assume the attacker has access to an oracle, which in this case amounts to a working scan-chain with the correct seed programmed in the LFSR.

To obtain enough algebraic expressions, our attack shifts in sequence of logic $0$s into the oracle scan chain obfuscated by the LFSR and captures the corresponding scan outputs $\boldsymbol{o}$. This is known as \emph{flushing} the scan chain\cite{8836102}. As we show below, choosing logic $0$s to scan in instead of random bits simplifies the algebraic expression of the scan output and corresponding final system of equations.

In particular, we can derive an algebraic representation of the secure scan chain. The matrix representation of the LFSR states reveals many properties  and can be derived from Equation~\ref{eq:lfsr1} as follows

\begin{equation}
\begin{pmatrix}
f^{t+1}_{0} \\
\vdots \\
f^{t+1}_{\lambda-2} \\
f^{t+1}_{\lambda-1} \\
\end{pmatrix} = 
\begin{pmatrix}
0 & 1 & \cdots & 0 \\
\vdots  & \vdots  & \ddots & \vdots  \\
0 & 0 & \cdots & 1 \\
c_0 & c_1 & \cdots & c_{\lambda-1} 
\end{pmatrix} 
\begin{pmatrix}
f^{t}_{0} \\
\vdots \\
f^{t}_{\lambda-2} \\
f^{t}_{\lambda-1} \\
\end{pmatrix} \\
\end{equation}
where, $t$ and $t+1$ represent the current and next cycle, respectively. We will refer to this transition matrix as $\boldsymbol{T}$. The state at any time step $t'$ can then be derived from the LFSR seed and $\boldsymbol{T}$ as follows
\begin{equation}
\begin{pmatrix}
f^{t'}_{0} \\
\vdots \\
f^{t'}_{\lambda-2} \\
f^{t'}_{\lambda-1} \\
\end{pmatrix} = 
\begin{pmatrix}
0 & 1 & \cdots & 0 \\
\vdots  & \vdots  & \ddots & \vdots  \\
0 & 0 & \cdots & 1 \\
c_0 & c_1 & \cdots & c_{\lambda-1} 
\end{pmatrix}^{t'}
\begin{pmatrix}
s_{0} \\
\vdots \\
s_{\lambda-2} \\
s_{\lambda-1} \\
\end{pmatrix} \\
\end{equation}

To simplify this representation, we use the matrix and vector forms as follows
\begin{equation}
\boldsymbol{f}^{t+1}=\boldsymbol{T}*\boldsymbol{f}^{t}
\end{equation}
\begin{equation}
\boldsymbol{f}^{t'}=\boldsymbol{T}^{t'}*\boldsymbol{s}
\label{eq:lfsrtransition}
\end{equation}

\noindent
Using Equation~\ref{eq:lfsrtransition}, we can symbolically represent the key input of any locking gate driven by the $i$th stage of the LFSR at time step $t'$:
\begin{equation}
f^{t'}_{i}=(\boldsymbol{T}^{t'}*\boldsymbol{s})[i]\\
\end{equation}

We observe that when logic $0$s go through the scan chain, they are simply XOR with keys $f^{t'}_{i}$. We can thus derive the symbolic expression for the expected values of the scan out signal. Let  $\boldsymbol{o_m}$ correspond to the scan output associated with the $m$th scan input. We then have

\begin{align}
    \boldsymbol{o_m}=&(\boldsymbol{T}^{m} \boldsymbol{s})[0] + (\boldsymbol{T}^{m+b} \boldsymbol{s})[1] + (\boldsymbol{T}^{m+2b} \boldsymbol{s})[2] \nonumber
    \\&+... + (\boldsymbol{T}^{m+(\lambda-1)b} \boldsymbol{s})[\lambda-1]
\end{align}

\noindent
By introducing an identity matrix $\boldsymbol{R}$ with shape $\lambda*\lambda$ and factoring out $\boldsymbol{s}$, we can further simplify this expression as follows

\begin{align}
\boldsymbol{o_m}=&[\boldsymbol{r_0} \boldsymbol{T}^{m}+\boldsymbol{r_1} \boldsymbol{T}^{m+b}+\boldsymbol{r_2} \boldsymbol{T}^{m+2b}\nonumber
\\&+... +\boldsymbol{r_{\lambda-1}}  \boldsymbol{T}^{m+(\lambda-1)b}]\boldsymbol{s}
\end{align}

\noindent
where $\boldsymbol{r_i}$ is the $i$th row of $\boldsymbol{R}$.
The size of the first term $\boldsymbol{a}=\boldsymbol{r_0} \boldsymbol{T}^{m}
+... +\boldsymbol{r_{\lambda-1}}  \boldsymbol{T}^{m+(\lambda-1)b}$ is $1*\lambda$.
Using the above $o_m$ symbolic equation repeatedly for $\lambda$ clock cycles and extracting their first term $a$, we can compose a system of linear equations in GF(2)
\begin{equation}
    \boldsymbol{As}=\boldsymbol{o}
    \label{eq:sle}
\end{equation}
where $\boldsymbol{A}$ consists of $\lambda\ \boldsymbol{a}$'s and $\boldsymbol{o}$ is the corresponding captured scan outputs.
Our attack completes by solving this system of equations in GF(2).

\subsection{Analysis of the Proposed Attack}
\label{sec:analysisofpa}
Since the system of linear equations in Eq. \ref{eq:sle} is based on the  physical structure of the circuit, it is guaranteed to be solvable.
If $\boldsymbol{A}$ is full-rank, 
the solution yields the unique 
secret seed vector $\boldsymbol{s}$. Otherwise, the solution yields 
a set of potential seed vectors
characterized by a particular solution of $\boldsymbol{As}=\boldsymbol{o}$ along with the null space of $\boldsymbol{A}$. 
More precisely, when the rank is $k$ less than $\lambda$, there are $2^k$ possible seeds.
These seeds can be used in further analysis, such as brute-force or SAT attacks, possibly in conjunction with attacking the combinational logic.

State of the art secure chains are protected by a shadow chain which prevents the scan chain from being influenced by the LFSR for first $\lambda$ clock cycles~\cite{Rahman2019DynamicallyOS}. Because the scan chain is longer than the LFSR, the first $o$ fully affected by the LFSR will be scanned out at cycle $N+1$. Interestingly, our attack can circumvent this defense by simply skipping the first $N$ scan outputs and collecting the next $\lambda$ scan outputs to compose the matrix $\boldsymbol{A}$.

% For the system of linear equations to have an exactly one solution, however, we need  $\lambda$ linear independent $\boldsymbol{a}$'s. 
%Our empirical studies suggest that a sufficient condition for the resulting system of equations to have precisely one solution is when
%$\boldsymbol{T}$ represents an LFSR with a maximal period, which is preferred for secure systems. However, even when the rank of $A$ is less than $\lambda$, the system of linear equations still constrains the secret seed. In particular, when the rank is $k$ less than $\lambda$, there are $2^k$  possible seeds characterized by the particular solution of $\boldsymbol{As}=\boldsymbol{o}$ along with the null space of $\boldsymbol{A}$.

%Since we are using symbolic algebra, the period of $\boldsymbol{T}$ only depends on the $c_i$, the short period cases from some seeds can be avoided. 
%For those LFSRs with inherently short period, we are at least able to reveal the relation between seed bits. Therefore, LFSRs are not equally vulnerable.

%In the defense systems, the LFSRs with long period are always desired \cite{WardlawLOT}, %the designs which can generate PN-sequences are most secure,
%whereas the short period LFSR is susceptible and should be avoided. As for the proposed attack, longer period LFSRs are more vulnerable, therefore our attack are effective on those dynamic defenses.

\subsection{Attack on MISR}
\begin{figure}[hbt!]
    \centering  
    \includegraphics[width=9.5cm]{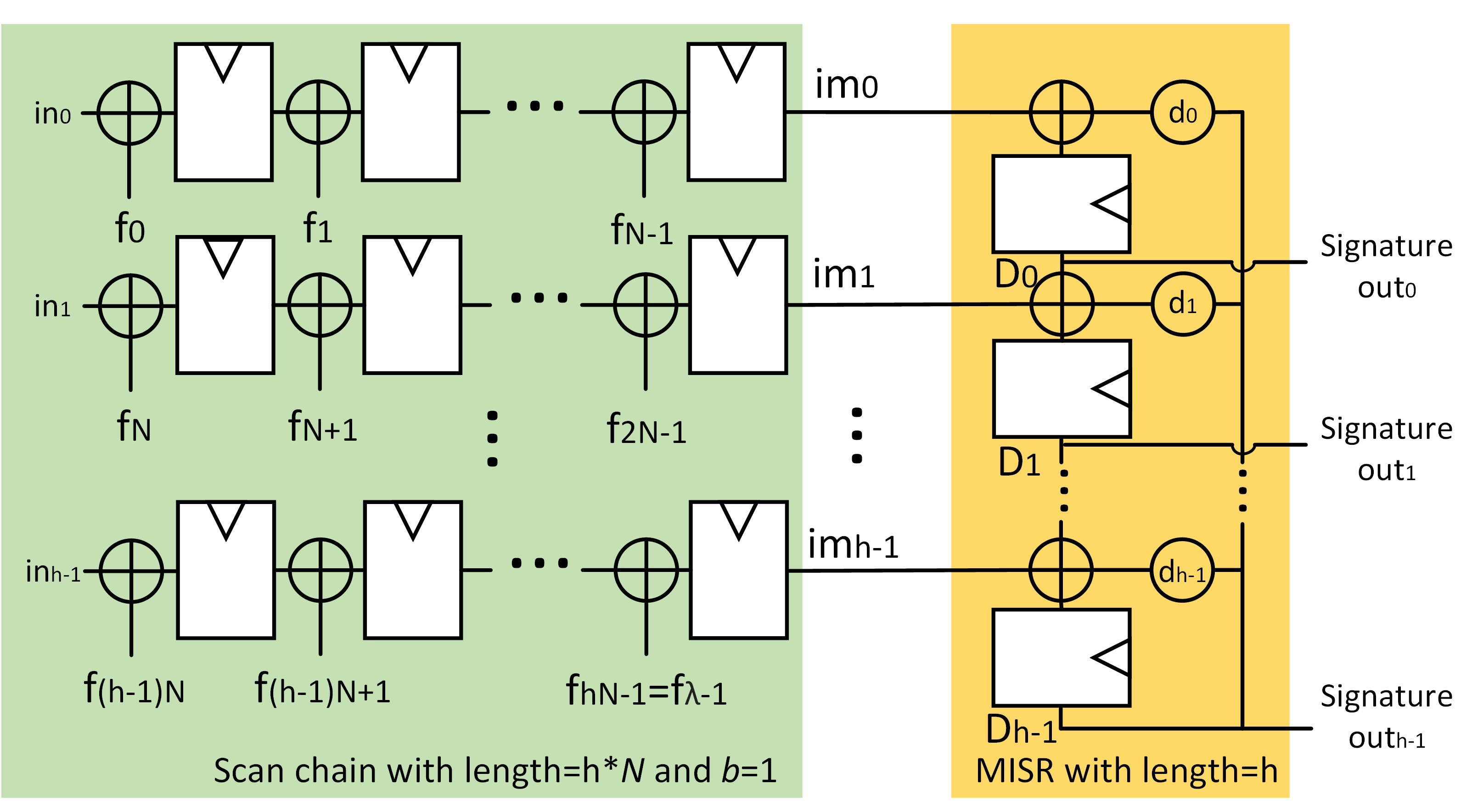}
    \caption{Structure of a secured scan chain with MISR}
    \label{fig: DOSwithMISR}
\end{figure}
Figure~\ref{fig: DOSwithMISR} shows the structure of dynamically secured scan chain with a MISR, where the length of every chain is $N$, the Boolean values $d_i$ define the structure of the MISR, and $D_i$ represent the internal Boolean state of MISR that is available for reading after every round of tests. We can observe that the MISR thwarts the direct access to scan outputs $im_i$. Importantly, the $h$ XOR gates in MISR are locking gates which corrupt the scan outputs $im$ and make attacks that demand direct access to scan outputs ineffective. Therefore, it is important to integrate the MISR into our algebraic model.

In our attack on scan chains with a MISR, we still shifts in sequence of logic 0s into scan chain, after $2N$ cycles, the MISR forms the signature outs $D^{2N}_i$ which we can read out.

In following equations, all superscripts denote the time stamps.
First of all, we derive the scan outputs $im_i$ from the 
LFSR keys $f$: 
\begin{equation}
im^{t}_i=\sum^{N-1}_{r=0} f^{t-N+r}_{r+iN}
\label{eq:im}
\end{equation}
where $im^{t}_i$ denotes the scan output of $i^{th}$ chain at cycle $t$, the sum is addition in GF(2), and all $f$ can be obtained using Equation~\ref{eq:lfsrtransition}. Then, we can derive $D^t_i$ as follows
\begin{equation}
D^{t}_0=im^{t-1}_0+d_0*D^{t-1}_{h-1}
\label{eq:D0}
\end{equation}
\begin{equation}
D^{t}_i=im^{t-1}_i+D^{t-1}_{i-1}+d_i*D^{t-1}_{h-1}\ \ for\ i>0
\label{eq:Di}
\end{equation}
where $D^{t}_i$ represents the internal values of the MISR stage $i$ at cycle $t$ and the initial $D_i^0$ are reset to 0. After $2N$ cycles, the signature outs are formed and available for reading:
\begin{equation}
signature\ out_i=D^{2N}_{i}
\label{eq:so}
\end{equation}
where every signature out is an equation with respect to seed bits, thus we obtain $h$ such equations in each round of test. 

%To maintain the dynamic property of the defense, 
We do not reset the LFSR but, as is typical, we reset the MISR at the beginning of every test sequence. Hence we require $\lambda/h=h*N/h=N$ tests, each involving $h$ equations with respect to seed bits, to obtain a sufficient number of equations to recover the secret seed. Similar to the analysis in Section~\ref{sec:analysisofpa}, a unique seed vector $s$ would be acquired in the case that these equations are full-rank, otherwise, we are able to acquire a  set  of potential  seed  vectors.

\section{Experimental Results}

\label{sec: exp} 
\subsection{Experiment Setup}

Our first experiment compares our algebraic attack to SAT-based attacks on scan chains and thus excludes a MISR.%We then consider the impact of a MISR.
\footnote{In practice, there often exists a bypass signal to circumvent the MISR. This analysis considers the case the attacker has  access to such a signal.}  
Both experiments demonstrate results for different key lengths. Since our attack isolates the scan chain, LFSR, and MISR, 
%we did not model the combinational logic driven by the scan chain in these experiments. %
there is no need to model the combinational logic driven by the scan chain. 
%THus, we  emphasize that the attack is equally effective with any benchmark circuit with a secured scan chain. 
%Moreover, the locking gates used in scan chain are XORs, because of the same effect for XOR and MUX as mentioned in section~\ref{bkg:dos}, the attack results is similar on MUX based counterparts.  
%
%For our experiments, 
We assume the key length $\lambda$ equals the scan chain length ($N$ without a MISR and $hN$ with a MISR), i.e., we set $b=1$.
%The LFSR used in all experiments have a period of $2^\lambda – 1$, which is considered to be the ideal defense. We excluded LFSR configurations $c$ that led to non-maximal periodicity.
The update of the LFSR is synchronized to the scan clock, which is also presumed to be the most secure defense. In addition, we assumed the existence of a shadow chain of length $\lambda$. %We thus skipped the first $\lambda$ scan outputs and collect the next $\lambda$ as $\boldsymbol{o}$ in Equation~\ref{eq:sle}.

We used MATLAB to generate the LFSR transition matrix $\boldsymbol{T}$, transition matrix of the secure scan chain $\boldsymbol{A}$ and MISR signature out recursively.
We then utilized the MATLAB function $\textit{gflineq}()$
and, when necessary, $\textit{gf2null}()$ to identify all the solutions 
%of $\boldsymbol{As}=\boldsymbol{o}$ 
over GF(2).
For each key length, we randomly chose 10 configuration vectors $c$, constrained to have $c_0=1$, made all $d_i=1$, and measured the average run-time including the generation of matrix $\boldsymbol{A}$ and $\boldsymbol{T}$ and the solving of the system of linear equations. 
All experiments were run on Intel i7-8700 CPU running at 3.20 GHz with 16-GB RAM.

%In  the  experiment  II, %placeholder, mention b, s and c

\subsection{Analysis of Basic Obfuscated Scan Chains}

\begin{figure}[hbt!]
    \centering  
    \includegraphics[width=6cm]{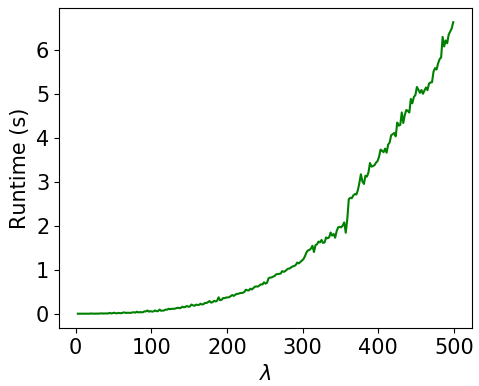}
    \caption{Average attack run-times vs. LFSR length $\lambda$}
    \label{fig: exp1}
\end{figure}
Figure \ref{fig: exp1} plots the average attack run-time on the defense without MISR as the number of key bits $\lambda$ ranging from 3 to 500. 
%and for each lambda we average over 10 random configurations.
Even with 500 key bits, the attack on average took less than 7 second. The run-time trend suggests the complexity of our attack scales as no more than a low-degree polynomial. This is expected because solving a system of linear equations has complexity no worse than $O(\lambda^3)$.
To further show the scalability of our proposed attack, we also tried $\lambda=1000$ and the attack took 66 seconds.

Interestingly, 87\% of the random configurations led to a unique seed, however, the average number of seeds is influenced by a few extreme cases and is $43.8$.
We further experimented with $\lambda=500$ and explored 1000 different random configurations of $c$.  The average number of seeds of 2.5 with the vast majority cases yielding a unique seed. We should emphasize however that for configurations where we could verify that the characteristic polynomial of the LFSR is primitive, a unique seed was always unveiled.

%\subsection{Monte Carlo Experiments}
%In most cases, the matrix $\boldsymbol{A}$ has sufficient rank and exactly one solution for seed.
%However, even if the rank of matrix $\boldsymbol{A}$ for some LFSR is $k$ less than $\lambda$, we still can narrow down the number of possible solutions to $2^k$.
%To evaluate the effectiveness of the proposed attack on different LFSRs, 
%Therefore, the proposed attack is able to identify the exact seed in most cases, and in a few cases, it can narrow down the search space to $2^2$.

\subsection{Analysis of Impact of MISRs}

\begin{figure}[bt!]
    \centering  
    \includegraphics[width=7cm]{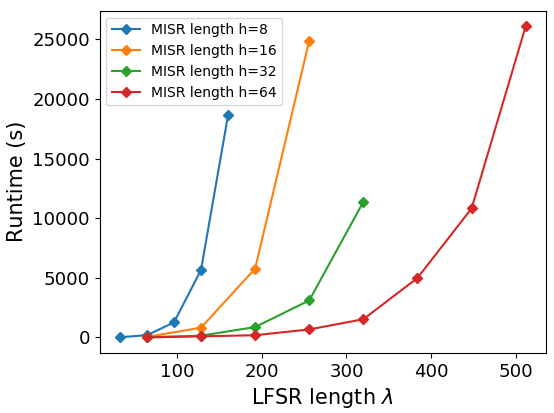}
    \caption{Average attack run-times with different size MISRs}
    \label{fig: exp2}
\end{figure}

Figure~\ref{fig: exp2} demonstrates the average attack run-times on the dynamically secured scan chain with different lengths of MISRs $h$ as a function of varying LFSR length $\lambda$ constrained by the relationship $\lambda=h*N$. %where $N$.
%is the length of each scan chain and $h$ varies from 8 to 64.
%the $\lambda$ has to be multiple of $h$, so we tested $\lambda=4h, 8h, 12h, 16h, 20h$ for $h=8$, $\lambda=4h, 8h, 12h, 16h$ for $h=16$, $\lambda=2h, 4h, 6h, 8h, 10h$ for $h=32$ and $\lambda=1h, 2h, 3h, 4h, 5h, 6h, 7h, 8h$ for $h=64$. We also tested greater $lambda$ and they  
The experiments with $\lambda>300$ 
timed-out after 8 hours for smaller values of $h$.
%$h=8, 16$, and $32$. 
This is because
with a MISR, we obtain only $h$ equations every test round (i.e., $2N$ cycles) compared to the case without a MISR which produces roughly one equation every cycle.  For practical MISR lengths that are typically greater than 16 ~\cite{7988042}, the attack run-time remains under 8 hours for LFSR lengths of under $250$. In all cases, the run-time is dominated by the computation of the various powers of the system matrix $T$.\footnote{Although not experimentally tested, we note that this computation can be parallelized across multiple processors by pre-computing increasingly larger powers of $T$ via iterative squaring.} %accelerated  the system matrix $T$.

%For the MISRs with small $h$, compared to the case without MISR which obtains an equation every cycle, the attack run-times are relatively large, but remains manageable. In the real cases, however, the length of MISR is usually greater than 16 or even more~\cite{7988042}, the attack run-time would be very small and practical.

\subsection{Comparison to Other Attacks}

State-of-the-art attacks on dynamically secured scan chains are based on SAT attacks \cite{8836102,9116197}. In particular, \cite{8836102} observed that the LFSR logic can be unrolled and combined with the associated combinational logic circuit and then attacked with SAT.  They tested their attack framework with various ISCAS benchmarks and demonstrated that even with 368 key bits they can successfully uncover the LFSR seed in less than one hour. 
However, their attack assumed the combinational logic was not logic locked, in contrast to what is advocated in \cite{Rahman2019DynamicallyOS}. This is an important limitation because several combinational logic techniques are known to be SAT-resistant \cite{yasin2017provably, 8395439} which would complicate this type of attack. Furthermore, the SAT-based attacks rely on the access to scan outputs. The presence of a MISR may restrict this access and should not be neglected. %However, it was paid little attention in previous SAT-based attacks.

In contrast, our proposed attack isolates the scan chain and in particular does not involve modeling or attacking the combinational logic and thus circumvents any effort to also unlock the combinational logic. Moreover, because it leverages the algebraic nature of the problem it can integrate the MISR into the model for attacking, and for the same defenses without MISR, it recovers the set of potential seeds orders of magnitude faster than equivalent SAT-based attacks .
% change experiments by adding runtime A and T
% scalability

\section{Conclusions and Discussion}

\label{sec:concl}
This paper presents a scalable GF(2) algebraic attack on scan chains that are obfuscated by dynamic keys generated by an LFSR. The experimental results demonstrate that the defenses with 500 key bits can be cracked in 7 seconds.
%
%To the best of our knowledge, this is the first algebraic analysis on the scan chain obfuscation techniques. 
The power of the proposed attack stems from the observation that all operations in the defensive circuitry can be modeled in GF(2). The results highlight that while SAT-attacks are powerful, algebraic attacks should not be overlooked as they can be significantly more efficient. 

The results lead to several ideas of improving secure scan chains to protect against such algebraic attacks. For example, obfuscating the structure of the LFSR or using non-linear LFSRs ~\cite{Hell2008, golomb1967shift} may 
make anticipating the scan output vectors more challenging. Studying whether such additional defenses can be circumvented with more sophisticated algebraic attacks becomes an important and interesting area of future work.

\small{\bibliographystyle{IEEEtran}
\bibliography{reference}}
\end{document}